\documentclass{elsart}

\usepackage{bm}       
\usepackage{amssymb}  
\usepackage{diagrams}

\makeatletter
\let\ea=\expandafter

\def\tvi{\vrule height12pt depth5pt width0pt}

\newbox\hyp \newbox\concl \newdimen\ruleWidth

\def\rule@nd{\rule@end}

\def\rule@ux#1&{
 \def\tmp{#1}
  \ifx\tmp\rule@nd
   \hskip-2em$
    \else\tmp\hskip2em\ea\rule@ux
     \fi}

\def\infer#1#2#3{\relax
 \setbox\hyp=\hbox{$\rule@ux#1&\rule@end&} \setbox\concl=\hbox{$#2$}
  \ifdim \wd\concl<\wd\hyp
   \ruleWidth=\wd\hyp
    \else\ruleWidth=\wd\concl
     \fi
     \advance\ruleWidth by 1cm
     $\vcenter{
       \vbox{\hbox to\ruleWidth{\hss\tvi \unhbox\hyp \hss}}
        \hrule height.7pt depth0pt width\ruleWidth
         \vbox{\hbox to\ruleWidth{\hss\tvi \unhbox\concl \hss}}}$
          \kern.1cm \hbox{#3}}
\makeatother

\newcommand\ie{\mbox{\textit{i.e.}}~}
\newcommand\etc{\mbox{\textrm{etc.}}}

\newcommand{\Saf}{\mathcal{S}}

\newcommand{\Set}{\mathbf{Set}}          
\newcommand{\Rel}{\mathbf{Rel}}          
\newcommand{\Int}{\mathbf{Int}}          
\newcommand{\PT}{\mathbf{PT}}          
\newcommand{\Pow}{\mathcal{P}}    
\newcommand{\Mulf}{\mathcal{M}_{\!f}}    
\newcommand\CoMon{\mathbf{CoMon}}
\newcommand\D{\mathop{{}\mathrm{D}}}
\newcommand\AC{\mathsf{AC}}
\newcommand\CtrAC{\mathsf{CtrAC}}
\newcommand\Bool{\mathbf{B}}

\newcommand\List{\mathsf{List}}
\newcommand\Skip{\mathsf{skip}}
\newcommand\Magic{\mathsf{magic}}
\newcommand\Abort{\mathsf{abort}}
\newcommand\LL{\mbox{\L}}
\newcommand\Perm{\mathfrak{S}}
\newcommand\Id{\mathsf{Id}}
\newcommand\I{{I}}

\newcommand{\Tensor}{\mathbin{\otimes}}
\newcommand{\Bottom}{\bot}
\newcommand{\One}{\mathbf{1}}
\newcommand{\Zero}{\mathbf{0}}
\newcommand{\LinearArrow}{\mathbin{\multimap}}
\newcommand{\Plus}{\mathbin{\oplus}}
\newcommand{\BigPlus}{\mathbin{\displaystyle{}\bigoplus{}}}
\newcommand{\With}{\mathbin{\&}}
\DeclareSymbolFont{fontexp}{OT1}{cmr}{bx}{n}
\DeclareMathSymbol{!}{\mathalpha}{fontexp}{`!}

\newcommand\sub{\subseteq}
\newcommand\morphism[1]{\ \stackrel{#1}{-->}\ }

\newcommand\be{\[\begin{array}[t]{llllllll}} \newcommand\ee{\end{array}\]}

\begin{document}

\begin{frontmatter}

\title{Interaction Systems \\
       and Linear Logic}
\subtitle{\small ?`A Different Games Semantics?}

\author{Pierre Hyvernat}
\ead{pierre.hyvernat@univ-savoie.fr}

\address{Laboratoire de math\'ematiques  \\
         universit\'e de Savoie          \\
         campus scientifique             \\
         73376 Le Bourget-du-Lac Cedex   \\
         France                            }

\begin{abstract}
  We define a model for linear logic based on two well-known ingredients:
  games and simulations. This model is interesting in the following respect:
  while it is obvious that the objects interpreting formulas are \emph{games}
  and that everything is developed with the intuition of \emph{interaction} in
  mind, the notion of morphism is very different from traditional morphisms in
  games semantics. In particular, we make no use of the notion of strategy!
  The resulting structure is very different from what is usually found in
  categories of games.

  We start by defining several constructions on those games and show, using
  elementary considerations,  that they enjoy the appropriate algebraic
  properties making this category a denotational model for intuitionistic
  linear logic. An interesting point is that the tensor product corresponds to
  a strongly \emph{synchronous} operation on games

  This category can also, using traditional translations, serve as a model for
  the simply typed $\lambda$-calculus.  We use some of the additional
  structure of the category to extend this to a model of the \emph{simply
  typed differential $\lambda$-calculus} of~\cite{diffLamb}.  Once this is
  done, we go a little further by constructing a reflexive object in this
  category, thus getting a concrete non-trivial model for the \emph{untyped}
  differential $\lambda$-calculus.

  We then show, using a highly non-constructive principle, that this category
  is in fact a model for full \emph{classical} linear logic ; and we finally
  have a brief look at the related notions of predicate transformers
  (\cite{denotPT}) and containers (\cite{cont}).
\end{abstract}


\end{frontmatter}

    %
%
%

%
%
%

\count255\catcode`@
\catcode`@=11
\chardef\mathlig@atcode\count255

\let\ea=\expandafter

\def\actively#1#2{\begingroup\uccode`\~=`#2\relax\uppercase{\endgroup#1~}}

\def\mathlig@gobble#1{\mathlig@next@cmd}

\def\mathlig@delim{\mathlig@delim}

\def\mathlig@defcs#1{\ea\def\csname#1\endcsname}

\def\mathlig@let@cs#1#2{\ea\let\ea#1\csname#2\endcsname}

\def\mathlig@appendcs#1#2{\ea\edef\csname#1\endcsname{\csname#1\endcsname#2}}

\def\mathlig#1#2{\mathlig@checklig#1\mathlig@end\mathlig@defcs{mathlig@back@#1}{#2}\ignorespaces}


\def\mathlig@checklig#1#2\mathlig@end{%
 \ea\ifx\csname mathlig@forw@#1\endcsname\relax
  \ea\mathchardef\csname mathlig@back@#1\endcsname=\mathcode`#1%
   \mathcode`#1"8000\actively\def#1{\csname mathlig@look@#1\endcsname}%
    \mathlig@dolig#1\mathlig@delim
     \fi
      \mathlig@checksuffix#1#2\mathlig@end}

\def\mathlig@checksuffix#1#2\mathlig@end{%
 \ifx\mathlig@delim#2\mathlig@delim\relax\else\mathlig@checksuffix@{#1}#2\mathlig@end\fi}

\def\mathlig@checksuffix@#1#2#3\mathlig@end{%
 \ea\ifx\csname mathlig@forw@#1#2\endcsname\relax\mathlig@dosuffix{#1}{#2}\fi
  \mathlig@checksuffix{#1#2}#3\mathlig@end}


\def\mathlig@dosuffix#1#2{%
\mathlig@appendcs{mathlig@toks@#1}{#2}%
\mathlig@dolig{#1}{#2}\mathlig@delim
}


\def\mathlig@dolig#1#2\mathlig@delim{%
 \mathlig@defcs{mathlig@look@#1#2}{%
 \mathlig@let@cs\mathlig@next{mathlig@forw@#1#2}\futurelet\mathlig@next@tok\mathlig@next}%
 \mathlig@defcs{mathlig@forw@#1#2}{%
  \mathlig@let@cs\mathlig@next{mathlig@back@#1#2}%
  \mathlig@let@cs\checker{mathlig@chck@#1#2}%
  \mathlig@let@cs\mathligtoks{mathlig@toks@#1#2}%
  \ea\ifx\ea\mathlig@delim\mathligtoks\mathlig@delim\relax\else
  \ea\checker\mathligtoks\mathlig@delim\fi
  \mathlig@next
 }%
 \mathlig@defcs{mathlig@toks@#1#2}{}%
 \mathlig@defcs{mathlig@chck@#1#2}##1##2\mathlig@delim{%
  \ifx\mathlig@next@tok##1%
   \mathlig@let@cs\mathlig@next@cmd{mathlig@look@#1#2##1}\let\mathlig@next\mathlig@gobble
  \fi 
  \ifx\mathlig@delim##2\mathlig@delim\relax\else
   \csname mathlig@chck@#1#2\endcsname##2\mathlig@delim
  \fi
 }%
%
 \ifx\mathlig@delim#2\mathlig@delim\else
  \mathlig@defcs{mathlig@back@#1#2}{\csname mathlig@back@#1\endcsname #2}%
 \fi
}%

\catcode`@\mathlig@atcode

\mathlig{|-}{\vdash}
\mathlig{|->}{\mapsto}
\mathlig{=>}{\Rightarrow}
\mathlig{<=>}{\Leftrightarrow}
\mathlig{->}{\rightarrow}
\mathlig{-->}{\longrightarrow}
\mathlig{-o}{\LinearArrow}
\mathlig{×}{\times}
\mathlig{||}{\mid}
\mathlig{·}{\cdot}
\mathlig{[[}{[\![} \mathlig{]]}{]\!]}


\section*{Introduction} 

Transition systems and simulation relations are well known tools in computer
science. More recent is the use of games to give models for different
programming languages \cite{AJM,HO}, or as an interesting tool for the study
of other programming notions \cite{ghica}.  We devise a denotational model of
linear logic based on those two ideas.  Basically, a formula will be
interpreted by an ``alternating transition system'' (called an
\emph{interaction system}) and a proof will be interpreted by a \emph{safety
property} for this interaction system.  Those concepts which were primarily
developed to model imperative programming and interfaces turned out to give a
rather interesting games model: a formula is interpreted by a game (the
interaction systems), and a proof by a witness that a non-loosing strategy
exists (the safety property). The notion of morphism corresponds to the notion
of simulation relation, a particular case of safety properties.

Part of the interest is that the notion of safety property is very simple.
They are only subsets of states in which the first player can remain, whatever
the second player does. In other words, from any state in the safety property,
the first player has an infinite strategy \emph{which never leaves the safety
property.} This is to be contrasted with traditional notions where morphisms
are \emph{functions} (usually depending on some subset of the history) giving
the strategy.

\smallbreak
The structure of safety properties is much richer than the structure of
proofs. In particular, safety properties are closed under arbitrary unions.
Since there is no notion of ``sum'' of proofs, this doesn't reflect a
logical property.  However, this is a feature rather than a bug: the
differential $\lambda$-calculus of Ehrhard and Regnier (\cite{diffLamb}) is an
extension of $\lambda$-calculus with a notion of differentiation and
non-deterministic sum. As we will show, interaction systems can interpret this
extra structure quite naturally.

Even better, this category enjoys such properties that we can, without much
difficulty, construct a \emph{reflexive object} allowing an interpretation of
\emph{untyped} differential $\lambda$-calculus. This reflexive object is
constructed using a fixpoint construction which is already available in the
category $\Rel$ of sets and relations between them.

The last thing we look at in this category is the object ``$\Bottom$''. As far
as interaction is concerned, this is one of the most boring objects. However,
it satisfies a very strong algebraic property: it is \emph{dualizing} and we
can thus interpret the whole of \emph{classical} linear logic. This was rather
unexpected and has a few surprising consequences (see
corollary~\ref{cor:ISSimple}). The main reason for this is that the principle
used to prove this fact (the contraposition of the axiom of choice) is
relatively counter-intuitive. It is also the only part of this work where
highly non-constructive principles are used, which explains why we separate
this from the rest.

\smallbreak
We conclude this paper by relating interaction systems with other notions,
namely the notion of \emph{predicate transformers} (\cite{denotPT}) and the
notion of \emph{containers} (\cite{cont}) and ``dependent containers''.

\section{Interaction Systems} 

The definition of interaction system we are using was developed primarily by
Peter Hancock and Anton Setzer. Their aim was to describe \emph{programming
interfaces} in dependent type theory (\cite{IOworld,IOworldBis}). The ability
to use dependent types makes it possible to add logical specifications to
usual specification. The result is a notion of formal interface describing:
\begin{itemize}
\item the way the programmer is allowed to use commands;
\item and the logical properties he can expect from those commands.
\end{itemize}
In practice, it is usually the case that the logical specification is ensured
\textit{a~posteriori}: a command might be legally used in a situation, even
though the logical specification prevents it

\smallbreak
Because of their definition however, interaction systems can be interpreted in
many different ways. In order to develop some intuitions, we prefer using a
``games'' interpretation: an interaction system describes the modalities of a
two persons game.

\subsection{The Category of Interaction Systems} 

Let's start with the raw definition:
\begin{defn}
  Let $S$ be a set (of \emph{states}); an \emph{interaction system} on $S$ is
  given by the following data:
  \begin{itemize}
    \item for each $s\in S$, a set $A(s)$ of \emph{possible actions;}
    \item for each $a\in A(s)$, a set $D(s,a)$ of \emph{possible reactions} to
      $a$;
    \item for each $d\in D(s,a)$, a \emph{new state} $n(s,a,d)\in S$.
  \end{itemize}
  We usually write $s[a/d]$ instead of $n(s,a,d)$.
  \smallbreak
  We use the letter $w$ for an arbitrary interaction system, and implicitly
  name its components $A$, $D$ and $n$. To resolve ambiguity, we sometimes
  use the notation $w.A$, $w.D$ and $w.n$ to denote the components of
  interaction system $w$. The set of states is usually implicit, and we write
  it $w.S$.
\end{defn}
Following standard practice within computer science, we distinguish the two
``characters'' by calling them the Angel (choosing actions, hence the $A$) and
the Demon (choosing reactions, hence the $D$).  Depending on the authors'
background, other names could be Player and Opponent, Eloise and Abelard,
Alice and Bob, Master and Slave, Client and Server, System and Environment,
alpha and beta, Arthur and Bertha, Left and Right \etc

\medbreak
As stated above, one of the original goals for interaction systems was to
represent programming interfaces. Here is for example the interface of a stack
of booleans:

\begin{tabular}{p{7cm}p{6cm}}
\begin{itemize}
  \item $S = \List(\Bool)$;
  \medbreak
  \item $\left\{\begin{array}[c]{l}
            A([]) = \{\mathsf{Push}(b) || b\in \Bool\}  \\
            A(\_) = \{\mathsf{Push}(b) || b\in \Bool\} \cup \{\mathsf{Pop}\}
         \end{array}\right.$
  \medbreak
  \item $ D(\_,\_) = \{\mathsf{Akn}\}$
  \medbreak
  \item $\left\{\begin{array}[c]{l}
           n(s,\mathsf{Push}(b)) = b::s  \\
           n(b::s,\mathsf{Pop}) = s
         \end{array}\right.$
\end{itemize}
&
\scriptsize\vskip0pt
The set of states $S$ represents the ``virtual'' internal states of an
implementation of stacks: lists (stacks) of booleans.

\medbreak

In a non-empty state, we can issue a ``$\mathsf{pop}$'' or a
``$\mathsf{push}(b)$'' command, but if the state is empty, we can only issue a
$\mathsf{push}$ command.

\medbreak

The responses are (in this case) trivial: we can only get an
``$\mathsf{Akn}$owledgment''.

\medbreak

The next state is defined by either adding a boolean in front of the list
($\mathsf{push}$) or removing the first element of the list ($\mathsf{pop}$).

\end{tabular}

What is still missing from this description is the ``side-effects'' part: it
doesn't say anywhere that a ``$\mathsf{pop}$'' command will return the first
element of the state.\footnote{It is possible to devise interaction systems
with such side-effects, but the theory of those is still to be
developed.} This is however much more precise than traditional interfaces
which are usually given by a collection of types. Compare with this poor
description of stacks:
\begin{tabular}{p{4cm}p{9cm}}
\begin{itemize}
  \item $\mathsf{Pop} : \Bool$
  \item $\mathsf{Push} : \Bool -> ()$
\end{itemize}
&
\scriptsize\vskip0pt
For any given stack, there are two commands: ``$\mathsf{Pop}$'' and
``$\mathsf{push}$'': the first one \emph{returns} a boolean and the second one
takes a single boolean argument and ``does something''.
\end{tabular}

Since we are aiming at a games semantics, we will rather use the following
interpretation:
\begin{itemize}
\item $S$ is a set of states describing the possible states of a
game;
\item for a state $s\in S$, the set $A(s)$ is the set of legal moves
in state $s$;
\item for such a move $a\in A(s)$, the set $D(s,a)$ is the set of countermoves
to $a$;
\item finally, $s[a/d]$ is just the new state of the game after $a$/$d$ has
been played.
\end{itemize}

\medbreak
Departing from the well established tradition of ``morphisms as strategies'',
we use the following notion of ``simulation relation'':
\begin{defn}
  If $w_1$ and $w_2$ are two interaction systems on $S_1$ and $S_2$
  respectively; a relation $r\subseteq S_1×S_2$ is called a \emph{simulation} if:
  \be
    (s_1,s_2)\in r \quad => \quad & \big(\forall a_1\in A_1(s_1)\big)\\
                                  & \big(\exists a_2\in A_2(s_2)\big)\\
                                  & \big(\forall d_2\in D_2(s_2,a_2)\big)\\
                                  & \big(\exists d_1\in D_1(s_1,a_1)\big)\\
                                  & \qquad \big(s_1[a_1/d_1],s_2[a_2/d_2]\big)\in r \ \mbox{.}
  \ee
\end{defn}
This definition is very similar to the usual definition of simulation
relations between labeled transition systems, but adds one layer of
quantifiers to deal with reactions. That $(s_1,s_2) \in r$ means that ``$s_2$
simulates $s_1$''.  By extension, if~$a_2$ is a witness to the first
existential quantifier, we say that ``$a_2$ simulates~$a_1$''.  Note that
because the left hand side would be vacuous, the empty relation is {always}
a simulation.

It is illuminating to look at the usual ``copycat strategy''
with this in mind: a simulation is just a generalization of what happens
there. Intuitively, a simulation from $w$ to $w'$ means that if the Angel
knows how to play in $w$, she can simulate, move after move, a play in $w'$.
(And vice and versa for the Demon.)

\smallbreak
To continue on the previous example, programming a stack interface amounts to
implementing the stack commands using a lower level interface (arrays and
pointer for examples). If we interpret the quantifiers constructively, this
amounts to providing a (constructive) proof that a non-empty relation is a
simulation from this lower level interaction system to stacks: for each of the
stacks commands, we need to provide a witness command in the low level world
in such a way as to guarantee simulation.  (See \cite{giovanni} and~\cite{PhD}
for a more detailed description of programming in terms of interaction
systems.)

\medbreak\noindent
Recall that the composition of two relations is given by:
\[
  (s_1,s_3) \in r_2·r_1  \quad<=>\quad  (\exists s_2)\ (s_1,s_2)\in r_1 \ \mbox{and}\
  (s_2,s_3)\in r_2
\]
It should be obvious that the composition of two simulations is a simulation
and that the equality relation is a simulation from any $w$ to itself.  Thus,
we can put:
\begin{defn}
  We call $\Int$ the category of interaction systems with simulations.
\end{defn}
Note that this category is locally small but that, for any given set of states
$S$, the collection of interaction systems on $S$ forms a proper class.  It is
possible to restrict to ``finitary'' interaction systems: those for which the
sets of actions and the sets of reactions are always finite. For our purposes,
it is impossible to restrict to finite sets of states (see the definition
of~$!w$) or to countable actions/reactions sets (see the definition
of~$w_1-ow_2$). Subtler considerations show that it is however possible to
restrict to sets of states of cardinalities $\aleph_0$ \emph{and} sets of
actions of cardinalities $2^{\aleph_0}$ and sets of reactions of cardinalities
$\aleph_0$. (See the proof of proposition~\ref{prop:PTInt} for a hint.)

\bigbreak
We have the following ``forgetful'' functor from $\Int$ to $\Rel$, the
category of sets and relations between them:
\begin{lem}
  The operation $w |-> |w| = w.S$ is a faithful functor from $\Int$ to~$\Rel$.
  Its action on morphisms is just the identity.

  This functor has a right adjoint $S|->\Magic(S)$ and a left adjoint
  $S|->\Abort(S)$ defined by
  {\scriptsize\hskip10mm (where $\{*\}$ denotes a singleton set)}
  \be
    \Magic(S).A(s)      &\quad=\quad& \{*\}     &\qquad&  \Abort(S).A(s)       &\quad=\quad& \emptyset \\
    \Magic(S).D(s,*)    &\quad=\quad& \emptyset &      &  \Abort(S).D(s,\_)    &\quad=\quad& \_        \\
    \Magic(S).n(s,*,\_) &\quad=\quad& \_        &      &  \Abort(S).n(s,\_,\_) &\quad=\quad& \_
  \ee
\end{lem}
In terms of games, $\Magic$ means ``the Demon resigns'' (he cannot answer any
move) while $\Abort$ means ``the Angel resigns'' (she cannot play).

This category enjoys a very strong algebraic property:
\begin{prop}  \label{prop:SupLatticeEnrich}
  The category $\Int$ is enriched over complete sup-lattices.
\end{prop}
Note that enrichment over sup-lattice is stronger than enrichment over
commutative monoids.
\begin{pf}
  Proving that an arbitrary unions of simulations in $\Int(w_1,w_2)$ is still a
  simulation in $\Int(w_1,w_2)$ is trivial; as well as showing that the empty
  relation is always a simulation.

  It thus only remains to show that composition commutes with unions, on the
  right and on the left. Since this is true in $\Rel$, it is also true in
  $\Int$!

\qed\end{pf}


\subsection{Notation} 

Let's recall some traditional notation.
\begin{itemize}

  \item An element of the indexed cartesian product $\prod_{a\in A} D(a)$ is
    given by a function $f$ taking any $a\in A$ to an $f(a)$ in $D(a)$. When the
    set $D(a)$ doesn't depend on $a$, it amounts to a function $f:A->D$.

  \item An element of the indexed disjoint sum $\sum_{a\in A} D(a)$ is given
    by a pair $(a,d)$ where $a\in A$ and $d\in D(a)$. When the set $D(a)$
    doesn't depend on $a$, this is simply the cartesian product $A×D$.

  \item We write $\List(S)$ for the set of ``lists'' over set $S$. A list is
    a tuple~$(s_1,s_2, \ldots s_n)$ of elements of $S$. The empty list
    is written $()$.

  \item The collection $\Mulf(S)$ of finite multisets over $S$ is the quotient
    of $\List(S)$ by permutations. We write $[s_1,\ldots s_n]$ for the
    equivalence class containing~$(s_1,\ldots s_n)$.
    We use ``$+$'' for the sum of multisets; it simply corresponds to
    concatenation on lists.
\end{itemize}

\subsection{Constructions} 

We now define, at the level of interaction systems, the connectives of linear
logic.  With those, making $\Int$ into a denotational model of intuitionistic
linear logic more or less amounts to showing that it is symmetric monoidal
closed, has finite products and coproducts and has a well behaved comonad.

\subsubsection{Constants.} 

A very simple, yet important interaction system is ``$\I$'', the interaction
system without interaction.
\begin{defn}
  Define $\I$ to be the interaction system on the singleton set
  $\{*\}$:
  \be
    A_\I(*)     &\quad=\quad& \{*\}\\
    D_\I(*,*)   &\quad=\quad& \{*\}\\
    n_\I(*,*,*) &\quad=\quad& \{*\} \ \hbox{.}
  \ee
  This interaction systems is also called ``\thinspace$\Skip$''.
\end{defn}
This is the perfect example of ``stable'' or ``constant'' game: no knowledge
is ever gained by any of the players. Since the interaction traces are
ultimately constant, this is sometimes considered a terminating system: both
sides have reached an agreement.

\bigbreak
One last interaction system of (theoretical) interest is given by the
interaction system on the empty set of states:
\begin{defn}
  Define $\Zero$ to be the unique interaction system on $\emptyset$.
\end{defn}
This interaction system is even more boring than $\I$: there are no states!
From a practical point of view, this system doesn't even exist. The following
is trivial:
\begin{lem}
  In $\Int$, the object $\Zero$ is a zero object: it is both initial and terminal.
\end{lem}

\subsubsection{Product and Coproduct} 

Since the forgetful functor $w|->|w|$ has a right  adjoint, we know it
commutes with colimits. As a result, we know that if the coproduct of $w_1$
and $w_2$ exists in~$\Int$, its set of states is isomorphic to the coproduct
of $S_1$ and $S_2$ in $\Rel$. We thus define:
\begin{defn}  \label{def:Plus}
  Suppose $w_1$ and $w_2$ are interaction systems on $S_1$ and $S_2$.
  Define the interaction system $w_1\Plus w_2$ on $S_1+S_2$ as
  follows:\footnote{Recall that $A+B = \{1\}{×}A \cup \{2\}{×}B$.}
  \be
    A_{w_1\Plus w_2}\big((i,s_i)\big)     &\quad=\quad & A_i(s_i)\\
    D_{w_1\Plus w_2}\big((i,s_i),a\big)   &\quad=\quad & D_i(s_i,a)\\
    n_{w_1\Plus w_2}\big((i,s_i),a,d\big) &\quad=\quad & (i,s_i[a/d])
  \ee
\end{defn}
In other words, the game $w_1\Plus w_2$ is simply a disjoint sum of $w_1$ and
$w_2$, and interaction takes place in only one of the games. Because we have
no initial state, there is no need to specify who, among the Angel or the
Demon, is making the choice of the game to use. With this in mind,
lemma~\ref{lem:ProdisCoprod} isn't very surprising.

\smallbreak
We have:
\begin{lem}
  The operation $\_\Plus\_$ is the coproduct in $\Int$.
\end{lem}
\begin{pf}
  We just ``lift'' the constructions from the category $\Rel$:

  \begin{itemize}
  \item injections: we put
  \be
    i_1 &\quad:\quad& \Int(w_1,w_1\Plus w_2) \\
        &\quad=\quad& \big\{\big(s_1,(1,s_1)\big) \ \mid\ s_1\in S_1\big\}
  \ee
  and similarly for $i_2 \in \Int(w_2,w_1\Plus w_2)$.

  \item copairing: suppose $r_1\in \Int(w_1,w)$ and $r_2\in \Int(w_2,w)$,
  define:
  \be
    [r_1,r_2] &\quad:\quad& \Int(w_1\Plus w_2,w) \\
              &\quad=\quad& \phantom{{}\cup{}} \big\{\big((1,s_1),s)\big) \ \mid\ (s_1,s)\in r_1\big\} \\
                      &&     {} \cup
                            \big\{\big((2,s_2),s)\big) \ \mid\ (s_2,s)\in r_2\big\}
  \ee
  \end{itemize}
  Checking that those constructions yield simulations is direct.

  Commutativity of the appropriate diagrams as well as universality can be
  lifted from~$\Rel$...

\qed\end{pf}

The next result is only surprising at first sight: the situation is similar
in~$\Rel$. It will however to be quite important in the sequel since we cannot
interpret the differential $\lambda$-calculus without it.
\begin{lem} \label{lem:ProdisCoprod}
  In $\Int$, $\Plus$ is also the product.
\end{lem}
\begin{pf}
  This is a direct consequence of commutative monoid enrichment
  (proposition~\ref{prop:SupLatticeEnrich}).\footnote{This is well known for
  abelian categories (see \cite[chap. 8]{MacLane}), but the existence of
  inverses is irrelevant.}

\qed\end{pf}
When dealing with linear logic, we use the usual symbol $\With$ when it
denotes a conjunction...

\subsubsection{Synchronous Product.} 

There is an obvious tensor construction reminiscent of the synchronous
product found in SCCS (synchronous calculus of communicating systems,
\cite{SCCS}):
\begin{defn}
  Suppose $w_1$ and $w_2$ are interaction systems on $S_1$ and $S_2$.
  Define the interaction system $w_1\Tensor w_2$ on $S_1×S_2$ as follows:
  \be
    A_{w_1\Tensor w_2}\big((s_1,s_2)\big)                     & \quad=\quad & A_1(s_1)×A_2(s_2)\\
    D_{w_1\Tensor w_2}\big((s_1,s_2),(a_1,a_2)\big)           & \quad=\quad & D_1(s_1,a_1)×D_2(s_2,a_2)\\
    n_{w_1\Tensor w_2}\big((s_1,s_2),(a_1,a_2),(d_1,d_2)\big) & \quad=\quad & \big(s_1[a_1/d_1],s_2[a_2/d_2]\big) \ \mbox{.}
  \ee
\end{defn}
This is a kind of \emph{lock-step synchronous parallel composition} of $w_1$
and $w_2$: the Angel and the Demon exchange pairs of actions/reactions.
In terms of games, the players simply play two games in parallel \emph{at the
same pace.}

For any sensible notion of morphism, $\I$ should be a neutral element for
this product. It is indeed the case, for the following reason: the components
of~$w\Tensor\I$ and $w$ are isomorphic by dropping the second (trivial)
coordinate:
\be
  w\Tensor \I                  &     &                        &                & w\\
  S×\{*\}                      &     &                        &\quad\simeq\quad& S\\
  A\big((s,*)\big)             &\ =\ & A(s) × \{*\}           &\quad\simeq\quad& A(s)\\
  D\big((s,*),(a,*)\big)       &\ =\ & D(s,a) × \{*\}         &\quad\simeq\quad& D(s,a)\\
  n\big((s,*),(a,*),(d,*)\big) &\ =\ & \big(s[a/d],*\big)     &\quad\simeq\quad& s[a/d]\\
\ee
This implies trivially that $\{( (s,*),s) || s\in S\}$ is an isomorphism.  For
similar reasons, this product is transitive and commutative.

\begin{lem}
  $\_\Tensor\_$ is a commutative tensor product in the category~$\Int$. Its
  action on morphisms is given by:
  \[
  \big((s_1,s'_1),(s_2,s'_2)\big) \in r\Tensor r'
    \quad<=>\quad
  \left\{\begin{array}[c]{ll}
                             &(s_1,s_2)\in r\\
           \mbox{\small and} &(s'_1,s'_2)\in r'
         \end{array}\right.
  \]
\end{lem}
Checking that $r\Tensor r'$ is a simulation is easy.

\medbreak
The notion of ``componentwise'' isomorphism is too fine for most purposes and
the notion of isomorphism inherited from simulations is coarser. In
particular, it is easy to find examples of isomorphic interaction systems in
$\Int$ where the actions/reactions sets are not isomorphic. For example, if
$w$ is an interaction system, let $\varepsilon_s$ be a choice function from
$\Pow^*\big(A(s)\big)$, the collection of non-empty subsets of $A(s)$ to $A(s)$;
and define $\widehat w$ with
\begin{itemize}
\item $\widehat A(s) = \Pow^*\big(A(s)\big)$;
\item $\widehat D(s,U) = D\big(s,\varepsilon_s(U)\big)$;
\item $\widehat n(s,U,d) = n\big(s,\varepsilon_s(U),d\big)$.
\end{itemize}
The systems $w$ and $\widehat w$ are isomorphic, but the sets $A(s)$ and
$\widehat A(s)$ have different cardinalities.

\subsubsection{Linear Arrow.} 

Any category with a zero object cannot be cartesian closed. We thus cannot
hope to model the simply typed $\lambda$-calculus inside $\Int$. One of the
points of linear logic is precisely to give logical status to a simpler kind
of structure: linear implication. We do not require our denotational model to
be cartesian closed but only symmetric monoidal closed w.r.t. to a tensor
which is generally not the cartesian product. This is the case for $\Int$, but
the definition of the functor $\_ -o \_$ is a little more involved:
\begin{defn}  \label{def:LinearArrow}
  If $w_1$ and $w_2$ are interaction systems on $S_1$ and $S_2$, define the
  interaction system $w_1-ow_2$ on $S_1×S_2$ as follows:
  \be
  A_{-o}\big((s_1,s_2)\big) \quad=\quad \displaystyle
                               \sum_{f\in A_1(s_1) -> A_2(S_2)}\ 
                               \prod_{a_1\in A_1(s_1)}
                               D_2\big(s_2,f(a_1)\big) -> D_1(s_1,a_1)\\
  D_{-o}\big((s_1,s_2),(f,G)\big) \quad=\quad \displaystyle\sum_{a_1\in A_1(s_1)} D_2\big(s_2,f(a_1)\big)\\
  n_{-o}\big((s_1,s_2),(f,G),(a_1,d_2)\big) \quad=\quad \big(s_1[a_1/G_{a_1}(d_2)]\,,\ s_2[f(a_1)/d_2]\big)
  \ \hbox{.}
  \ee
\end{defn}
It may seem difficult to get some intuition about this interaction system; but
it is \textit{a posteriori} quite natural. Let's unfold this definition with
simulations in mind:
\begin{itemize}
  \item An action in state $(s_1,s_2)$ is given by a pair consisting of:
    \begin{enumerate}
      \item[\scriptsize(1)] a function $f$ (the index for the element of the disjoint sum)
      translating actions from $s_1$ into actions from $s_2$;
      \item[\scriptsize(2)] for any action $a_1$, a function $G_{a_1}$ translating reactions
        to $f(a_1)$ into reactions to $a_1$.
    \end{enumerate}
  \item A reaction to such a ``one step translating mechanism'' is given by:
    \begin{enumerate}
      \item[\scriptsize(2)] an action $a_1$ in $A_1(s_1)$ (which we want to simulate);
      \item[\scriptsize(1)] and a reaction $d_2$ in $D_2(s_2,f(a_1))$ (which we want to
        translate back).
    \end{enumerate}
  \item Given such a reaction, we can simulate $a_1$ by $a_2=f(a_1)\in
    A_2(s_2)$; and translate back $d_2$ into $d_1=G_{a_1}(d_2)\in
    D_1(s_1,a_1)$. The next state is just the pair of states $s_1[a_1/d_1]$
    and $s_2[a_2/d_2]$.
\end{itemize}
In essence, the Angel translates what the Demon gives her.

This connective is indeed a ``linear arrow'':
\begin{prop}  \label{prop:adjoint}
  In $\Int$, $\_\Tensor\_$ is left adjoint to $\_-o\_$: there is an
  isomorphism
  \[
    \Int\big( w_1\Tensor w_2\,,\,w_3\big)
    \quad\simeq\quad
    \Int\big( w_1\,,\,w_2-ow_3\big)
    \ \mbox{,}
  \]
  natural in $w_1$, $w_2$ and $w_3$ 
\end{prop}

\begin{pf}
  The proof is not really difficult. First notice that the axiom of choice can
  be written as
  \be
  \AC :\quad &
  \big(\forall a\in A\big)\big(\exists d\in D(a)\big)\ \varphi(a,d) \\
             &<=>\\
  &\big(\exists f\in\prod_{a\in A} D(a)\big)\big(\forall a\in A\big)\ \varphi\big(a,f(a)\big)
  \ \mbox{.}
  \ee
  When the domain $D(a)$ for the existential quantifier doesn't depend on
  $a\in A$, we can simplify it into:
  \be
  \AC :\quad &
  \big(\forall a\in A\big)\big(\exists d\in D\big)\ \varphi(a,d) \\
   & <=> \\
  & \big(\exists f\in A -> D\big)\big(\forall a\in A\big)\ \varphi\big(a,f(a)\big)
  \ \mbox{.}
  \ee

  \smallbreak
  We will use $\AC$ to shuffle quantifiers and complexify the domains of
  quantification. This will transform the condition defining a simulation from
  $w_1\Tensor w_2$ to $w_3$ into the condition defining a simulation from $w_1$
  to $w_2-ow_3$.

  In the sequel, the part of the formula being manipulated will be written in
  bold.  That $r$ is a simulation from $w_1\Tensor w_2$ to $w_3$ takes the
  form\footnote{modulo associativity $(S_1×S_2)×S_3 \simeq S_1×(S_2×S_3)\simeq
  S_1×S_2×S_3$.}
  \be
    (s_1,s_2,s_3)\in r &=>& \big(\forall a_1\in A_1(s_1)\big)\bm{\big(\forall a_2\in A_2(s_2)\big)}\\
                       &  & \bm{\big(\exists a_3\in A_3(s_3)\big)}\\
                       &  & \big(\forall d_3\in D_3(s_3,\bm{a_3})\big)\\
                       &  & \big(\exists d_1\in D_1(s_1,a_1)\big)\big(\exists d_2\in D_2(s_2,a_2)\big)\\
                       &  & \quad \big(s_1[a_1/d_1],s_2[a_2/d_2],s_3[\bm{a_3}/d_3]\big) \in r
  \ \mbox{.}
  \ee
  Using $\AC$ on $\forall a_2\exists a_3$, we obtain:
  \be
  (s_1,s_2,s_3)\in r &=>& \big(\forall a_1\in A_1(s_1)\big)\\
                     &  & \big(\exists f\in A_2(s_2) -> A_3(s_3)\big)\\
                     &  & \big(\forall a_2\in A_2(s_2)\big)\bm{\big(\forall d_3\in D_3(s_3,f(a_2))\big)}\\
                     &  & \big(\exists d_1\in D_1(s_1,a_1)\big)\bm{\big(\exists d_2\in D_2(s_2,a_2)\big)}\\
                     &  & \quad \big(s_1[a_1/d_1],s_2[a_2/\bm{d_2}],s_3[f(a_2)/d_3]\big) \in r
  \ \mbox{.}
  \ee
  We can now apply $\AC$ on $\forall d_3\exists d_2$:
  \be
  (s_1,s_2,s_3)\in r &=>& \big(\forall a_1\in A_1(s_1)\big)\\
                     &  & \big(\exists f\in A_2(s_2) -> A_3(s_3)\big)\\
                     &  & \bm{\big(\forall a_2\in A_2(s_2)\big)}\\
                     &  & \bm{\big(\exists g\in D_3(s_3,f(a_2)) \to D_2(s_2,a_2)\big)}\\
                     &  & \big(\forall d_3\in D_3(s_3,f(a_2))\big)\\
                     &  & \big(\exists d_1\in D_1(s_1,d_1)\big)\\
                     &  & \quad \big(s_1[a_1/d_1],s_2[a_2/\bm{g}(d_3)],s_3[f(a_2)/d_3]\big) \in r
  \ee
  and apply $\AC$ one more time on $\forall a_2\exists g$ to obtain:
  \be
  (s_1,s_2,s_3)\in r &=>& \big(\forall a_1\in A_1(s_1)\big)\\
                     &  & \big(\exists f\in A_2(s_2) -> A_3(s_3)\big)\\
                     &  & \big(\exists G\in \prod_{a_2\in A_2(s_2)} D_3(s_3,f(a_2)) -> D_2(s_2,a_2)\big)\\
                     &  & \big(\forall a_2\in A_2(s_2)\big)
                          \big(\forall d_3\in D_3(s_3,f(a_2))\big)\\
                     &  & \big(\exists d_1\in D_1(s_1,d_1)\big)\\
                     &  & \quad \big(s_1[a_1/d_1],s_2[a_2/G_{a_2}(d_3)],s_3[f(a_2)/d_3]\big) \in r
  \ee
  which is equivalent to
  \be
  (s_1,s_2,s_3)\in r &=>& \big(\forall a_1\in A_1(s_1)\big)\\
                     &  & \left(\exists (f,G) \in \begin{array}[c]{l}
                                                   \sum_{f\in A_2(s_2)-> A_3(s_3)}\\
                                                   \prod_{a_2\in A_2(s_2)}  D_3(s_3,f(a_2)) ->  D_2(s_2,a_2)
                                                 \end{array}\right)\\
                     &  & \big(\forall (a_2,d_3) \in \sum_{A_2(s_2)} D_3(s_3,f(a_2))\big)\\
                     &  & \big(\exists d_1\in D_1(s_1,d_1)\big)\\
                     &  & \quad \big(s_1[a_1/d_1],s_2[a_2/G_{a_2}(d_3)],s_3[f(a_2)/d_3]\big) \in r
  \ \mbox{.}
  \ee
  By definition, this means that $r$ is a simulation from $w_1$ to $w_2-ow_3$.

  Naturality is trivial: it corresponds to naturality of associativity in
  $\Rel$.
\qed\end{pf}
In particular, proposition~\ref{prop:adjoint} implies that
\[
  \Int(w_1,w_2) \quad\simeq\quad \Int(\I,w_1-ow_2)
  \ \mbox{.}
\]
We call a simulation from $\I$ to $w$ a \emph{safety property} for $w$.
\begin{lem}[with Def.]
  A subset $x\sub S$ is a simulation from $\I$ to $w$ iff\footnote{This is
  well defined since $\Pow(\{*\}{×}S)\simeq\Pow(S)$.}
  \[
  s\in x  \quad=>\quad \big(\exists a\in A(s)\big)\big(\forall d\in D(s,a)\big)\ s[a/d]\in x
  \ \mbox{.}
  \]
  We write $\Saf(w)$ for the collection of such subsets, and we call such an
  $x$ a \emph{safety property for $w$.}

  Finally, we have
  \[
    \Int(w_1,w_2) \quad=\quad \Saf(w_1 -o w_2)\ \mbox{.}
  \]
\end{lem}
The analogy with traditional notions of morphisms as strategies is rather
subtle. A safety property~$x$ satisfies the property
\begin{quote}\sl
  if interaction is started from a state in $x$, then the Angel has a move
  that guarantees that the next state will also be in $x$ (provided the Demon
  does answer).
\end{quote}
This is a safety property in the sense that it guarantees that ``nothing bad
happens''. (As opposed to liveness properties, which ensure that ``something good happens''...)
In particular, this means that the Angel has an infinite strategy from any
state in~$x$: she can always find a move to play. The choice of those moves is
irrelevant to the notion of safety property: we only know they exist.
In particular, such a move needs not be unique.

\medbreak
As special case, let's look at the definition of linear negation.  The
orthogonal~$w^\Bottom$ of $w$ is defined as usual as the interaction system $w
-o \Bottom$. For intuitionistic linear logic, any object can formally be used
as $\Bottom$, but anticipating on proposition~\ref{prop:dual}, we use
$\Bottom = \I$. We have:
\be  \label{def:Bottom}
  A^\Bottom \big((s,*)\big) \quad=\quad \displaystyle
                              \sum_{f\in A(s) -> \{*\}}\ \prod_{a\in A(s)}
                              \{*\} -> D(s,a)\\
  D^\Bottom \big((s,*),(f,G)\big) \quad=\quad \displaystyle \sum_{a\in A(s)} \{*\}\\
  n^\Bottom \big((s,*),(f,G),(a,*)\big) \quad=\quad \big(s[a/G_{a_1}(*)],*\big)
\ee
which, after simplification, is equivalent to
\be
  A^\Bottom (s)     &\quad=\quad& \displaystyle\prod_{a\in A(s)} D(s,a)\\
  D^\Bottom (s,f)   &\quad=\quad& A(s)\\
  n^\Bottom (s,f,a) &\quad=\quad& s[a/f(a)] \ \hbox{.}
\ee
One important point to notice is that with this definition, the set of states
of~$w^\Bottom$ is the same as the set of states of $w$. In particular, the
canonical morphism in $\Int(w,w^{\Bottom\Bottom})$ will be given by the
equality relation.

The definition looks complex but can be interpreted in a very traditional way:
\emph{negation interchanges the two players.} In our context, it is not
possible to simply interchange actions and reactions since reactions
\emph{depend} on a particular action. We have to do the following:
\begin{itemize}
\item an action in $A^\Bottom(s)$ is a \emph{conditional} reaction; or a
\emph{one move strategy} to react to any action;
\item a reaction in $D^\Bottom(s,f)$ is just an action in $w$;
\item the new state after ``action'' $f$ and ``reaction'' $a$ is just the
state obtained after playing $a$ followed by $f(a)$.
\end{itemize}
The ``polarities'' of moves and coutermoves is interchanged, and it does swap
the players in the sense that it transforms Angel strategies into Demon
strategies and vice and versa. (See~\cite{PhD} for more details.)

The dual $w^\Bottom$ enjoys a surprising property: the set of possible
reactions to a particular action \emph{doesn't depend on the particular
action!}

\subsubsection{Multithreading.} 

We now come to the last connective needed to interpret intuitionistic linear
logic.  Its computational interpretation is related to the notion of
\emph{multithreading,} \ie the possibility to run several instances of a
program in parallel. In our case, it corresponds to the ability to play
several synchronous instances of the same game in parallel. Let's start by
defining synchronous multithreading in the most obvious way:
\begin{defn}
  If $w$ is an interaction system on $S$, define $\LL(w)$, the multithreaded
  version of $w$ to be the interaction system on $\List(S)$ with:
  \be
    \LL.A\big((s_1,\ldots,s_n)\big)                                   &=& A(s_1)×\ldots×A(s_n)\\
    \LL.D\big((s_1,\ldots,s_n),(a_1,\ldots,a_n)\big)                  &=& D(s_1,a_1)×\ldots×D(s_n,d_n)\\
    \LL.n\big((s_1,\ldots,s_n),(a_1,\ldots,a_n),(d_1,\ldots,d_n)\big) &=& \big(s_1[a_1/d_1],\ldots,s_n[a_n/d_n]\big) \ \hbox{.}
  \ee
\end{defn}
This interaction system is just the sum of all ``$n$-ary'' versions of the
synchronous product.
\begin{lem}
  The operator $w |-> \LL(w)$ can be extended to a functor from $\Int$ to
  $\Int$.
\end{lem}
To get the abstract properties we want, we need to quotient multithreading
by permutations. Just like multisets are lists modulo permutations, so is~$!w$
the multithreaded $\LL(w)$ modulo permutations.  This definition is possible
because~$\LL(w)$ is ``compatible'' with permutations: if $\sigma$ is a
permutation, we have
\be
 \sigma·\big((s_1,\ldots s_n)\big[(a_1,\ldots a_n)/(d_1,\ldots d_n)\big]\big)\\
  \qquad=\\
  \big(\sigma·(s_1,\ldots s_n)\big)[\sigma·(a_1,\ldots a_n)/\sigma·(d_1,\ldots d_n)]
  \ \hbox{.}
\ee
The final definition is:
\begin{defn}
  If $w$ is an interaction system on $S$, define $!w$ to be
  the following interaction system on $\Mulf(S)$:
  \be
  !A(\mu) &\quad=\quad& \sum_{\overline{s}\in\mu} \LL.A(\overline{s})\\
  !D\big(\mu,(\overline{s},\overline{a})\big) &\quad=\quad& \LL.D(\overline{s},\overline{a})\\
  !n\big(\mu,(\overline{s},\overline{a}),\overline{d}\big) &\quad=\quad&
          [\LL.n(\overline{s},\overline{a},\overline{d})] \ \hbox{.}
  \ee
  {\footnotesize(Note that as an element of $\mu$, $\overline{s}$ is just a
  specific order for the element of $\mu$.)}

\end{defn}
Unfolded, it gives:
\begin{itemize}
  \item an action in state $\mu$ is given by an element $\overline{s}$ (a
    list) of $\mu$ (a multiset, \ie an equivalence class) together with an
    element $\overline{a}$ in $\LL.A(\overline{s})$ (a list of actions);
  \item a reaction is given by a list of reactions $\overline{d}$ in
    $\LL.D(\overline{s},\overline{a})$;
  \item the next state is the equivalence class containing the list
    $\overline{s}[\overline{a}/\overline{d}]$ (the orbit
    of~$\overline{s}[\overline{a}/\overline{d}]$ under the action of the group
    of permutations).
\end{itemize}

\begin{lem} \label{lem:OfCourseFunctor}
  This operation $w |-> !w$ can be extended to a functor from $\Int$
  to~$\Int$.
  Moreover, we have the following bisimulation (which is not an isomorphism)
  \begin{equation} \label{eqFromLwToOfCoursew}
    \LL(w)
    \quad
    \mathop{\stackrel{\displaystyle\rightarrow}{\leftarrow}}\limits^\sigma_p
    \quad
    !w
  \end{equation}
  where $\sigma$ is just membership of a list in a multiset (equivalence class
  of lists) and $p$ its converse.
\end{lem}

This operation enjoys a very strong algebraic property:
\begin{prop} \label{prop:FreeComonoid}
  $!w$ is the free $\Tensor$-comonoid generated by $w$.
\end{prop}
Note that because $!w$ ``is'' $\BigPlus_{n\geq 0} w^{\Tensor n}/\Perm_n$, this
is not very surprising.
\begin{pf}
Quite a lot can be deduced from the same property of $\Rel$, but we will
look at some details.

Let's start by looking at the $\Tensor$-comonoid structure of $!w$: the counit
and comultiplication are given by
\be
  e &\ \in\ & \Int(!w,\I) &\quad\mbox{and}\quad&  m &\ \in\ & \Int(!w,!w\Tensor!w)\\
    &\ =\   & \{([],*)\}    &                    &    &\ =\   & \{\big(\mu+\nu,(\mu,\nu)\big)\ \mid\ \mu,\nu\in\Mulf(S)\}
\ee
We need to show that for any interaction system $w$ and $\Tensor$-comonoid
$w_c$, there is a natural isomorphism
\[
    \CoMon(\Int,\Tensor)(w_c,!w) \quad\simeq\quad \Int(w_c,w)
    \ \mbox{.}
\]
Going from left to right is easy:
\be
    \CoMon(\Int,\Tensor)(w_c,!w) &\quad->\quad& \Int(w_c,w) \cr
    r                            &\quad|->\quad& \{(s_c,s)\ \mid\ (s_c,[s]) \in r\}
    \ \mbox{.}
\ee
Checking that this operation is well-defined (it sends a comonoid morphism to
a simulation) is direct.

\smallbreak
The other direction is more interesting. Let $w_c$ be a commutative comonoid.
This means we are given $e_c\in\Int(w_c,\I)$ and $m_c\in\Int(w_c,w_c\Tensor
w_c)$, satisfying additional commutativity and associativity conditions.

Suppose $r$ is a simulation from $w_c$ to $w$. This is a relation with no
condition about the comonoid structure of $w_c$. We construct a relation
from~$w_c$ to~$!w$ in the following way:
\begin{itemize}
\item we start by extending comultiplication $m_c$ to
$\overline m_c:\Int\big(w_c,\LL(w_c)\big)$;
\item we then compose that with $\LL(r):\Int\big(\LL(w_c),\LL(w)\big)$;
\item and finally compose that with $\sigma:\Int\big(\LL(w),!w\big)$,
see~$(\ref{eqFromLwToOfCoursew})$ in lemma~\ref{lem:OfCourseFunctor}.
\end{itemize}
We then check that this simulation respects the comonoid structures of $w_c$
and~$!w$.

Define $\overline m_c \sub S_c×\List(S_c)$ by the following clauses:
(inductive definition)
  \be
    \big(s,()\big) \in \overline m_c              &\quad\hbox{iff}\quad& s\in e_c\cr
    \big(s,s'\big) \in \overline m_c              &\quad\hbox{iff}\quad& s=s'    \cr
    \big(s,(s_1,\dots,s_n)\big) \in \overline m_c &\quad\hbox{iff}\quad&
        \big(s,(s_1,s')\big) \in m_c \land \big(s',(s_2,\dots,s_n)\big) \in\overline m_c \cr
        &&\hbox{for some $s'\in S_c$}
  \ \mbox{.}
  \ee
Using the fact that $e_c$ and $m_c$ are simulations, we can easily show (by
induction) that $\overline m_c$ is a simulation from $w_c$ to $\LL(w_c)$.

Moreover, we have:
\begin{equation} \label{eqComonoidTrans}
\begin{array}{llllll}
\big(s_c,(s_{c,1},\dots,s_{c,n+m})\big) \in \overline{m_c} \cr
\quad<=>\quad \cr
(\exists s_c^1,s_c^2 \in S_c)\ \big(s_c,(s_c^1,s_c^2)\big) \in m_c
                   \vtop{\halign{&$#$\hfill\cr
                         {}\land \big(s_c^1,(s_{c,1},\dots,s_{c,n})\big) \in \overline m_c \cr
                         {}\land \big(s_c^2,(s_{c,n+1},\dots,s_{c,n+m})\big) \in \overline m_c
                                     \phantom{\Big(}
                         \cr}}
\end{array}
\end{equation}
by transitivity and
\begin{equation} \label{eqComonoidCom}
\begin{array}{llllll}
\big(s_c,(s_{c,1},\dots,s_{c,i},s_{c,i+1},\dots,s_{c,n})\big) \in \overline{m_c} \cr
\quad<=>\quad \cr
\big(s_c,(s_{c,1},\dots,s_{c,i+1},s_{c,i},\dots,s_{c,n})\big) \in \overline{m_c} \cr
\end{array}
\end{equation}
by commutativity.

We know that $\widetilde r = \sigma·\LL(r)·\overline m_c$ is a simulation
from~$w_c$ to~$!w$.  We need to check that this simulation respects the
comonoid structures of $w_c$ and $!w$, \ie that both
\diagram[height=2em,width=4em,tight]
 w_c                   & \rTo^{\widetilde r}       & !w &\qquad&
             w_c                 & \rTo^c\mskip5mu & w_c\Tensor w_c                         \\
 &\rdTo_{e_c}    & \dTo_{e}        &\quad\hbox{and}\quad&
             \dTo^{\widetilde r} &                   & \dTo_{\widetilde r\Tensor\widetilde r} \\
                       &              &\One &\qquad&
             !w                  & \rTo_{m_c}\mskip5mu   & !w\Tensor!w                          \\
\enddiagram
are commutative.
The first diagram is easily shown to be commutative.  For the second one:
suppose $\big(s_c,[s_1,\dots,s_n],[s_{n+1},\dots,s_{n+m}]\big)\in m·\widetilde
r$.  This is equivalent to saying that there are $s_{c,1},\dots,s_{c,n+m}$ in
$S_c$ s.t.
\begin{itemize}
\item $(s_{c,i},s_i) \in r$ for all $i=1,\dots,n+m$
\item and $\big(s_c,(s_{c,1},\dots,s_{c,n+m})\big) \in \overline m_c$.
\end{itemize}
That $\big(s_c,[s_1,\dots,s_n],[s_{n+1},\dots,s_{n+m}]\big)$ is in $\widetilde
r\Tensor\widetilde r · c$ means that there are~$s_c^1$ and~$s_c^2$ in~$S_c$
s.t.
\begin{itemize}
\item $\big(s_c,(s_c^1,s_c^2)\big) \in m_c$
\item and $\big(s_c^1,[s_{1},\dots,s_{n}]\big) \in \widetilde r$
and~$\big(s_c^2,[s_{n+1},\dots,s_{n+m}]\big) \in \widetilde r$,
\end{itemize}
\ie there are $s_c^1$ and $s_c^2$ in $S_c$, and
$s_{c,1},\dots,s_{c,n},s_{c,n+1},\dots,s_{c,n+m}$ in $S_c$ s.t.
\begin{itemize}
\item $\big(s_c,(s_c^1,s_c^2)\big) \in m_c$
\item $\big(s_c^1,(s_{c,1},\dots,s_{c,n})\big) \in \overline m_c$
\item $\big(s_c^2,(s_{c,n+1},\dots,s_{c,n+m})\big) \in \overline m_c$
\item and $(s_i,s_{c,i})\in r$ for all $i=1,\dots,n+m$.
\end{itemize}
By using $(\ref{eqComonoidTrans})$ and $(\ref{eqComonoidCom})$, it is trivial
to show that the two conditions are equivalent. This proves that the second
diagram is commutative.

It only remains to show that the two operations defined are inverse of each
other. This is not difficult.
\qed\end{pf}


\section{Interpreting Linear Logic} 
\label{sec:ILL}

We now have all the necessary ingredients to construct a denotational model
for intuitionistic linear logic. The details of categorical models for linear
logic are quite intricate, and there are several notions, not all of which are
equivalent. We refer to the survey~\cite{CatLLMellies} and the references given there.

In the case of $\Int$, the situation is however quite simple:
proposition~\ref{prop:FreeComonoid} makes~$\Int$ into a ``Lafont category''
(see~\cite{PhDLafont}).
\begin{cor}
  With the construction defined in the previous sections, $\Int$ is a Lafont category.
  In particular, ``\/$!\_$'' is a comonad; and we have for any $w_1$
  and $w_2$, we have the following natural isomorphism:
  \[
    !(w_1 \With w_2) \quad\simeq\quad !w_1 \Tensor !w_2
    \ \mbox{.}
  \]
  {\scriptsize Recall that since the product and coproduct coincide, $\With$
  is the same as $\Plus$.}
\end{cor}
A direct proof of the fundamental isomorphism is easy: there is a
``componentwise'' isomorphism between the interaction systems $!(w_1\With
w_2)$ and $!w_1\Tensor!w_2$.

Lafont categories were used to give a semantics to linear logic and were
latter subsumed by Seely categories and linear categories.

It is thus possible to give a semantics to formulas and proofs in the usual
way. We write~$F$ for the interpretation of a formula~$F$ (no confusion
arises) and~$[[\pi]]$ for the interpretation of a proof~$\pi$.
\begin{prop} \label{prop:red_inv}
  For all proof $\pi_1$ and $\pi_2$ of the same sequent, if $\pi_1$ and
  $\pi_2$ have the same cut-free normal form, then $[[\pi_1]]=[[\pi_2]]$.
\end{prop}
Moreover, since the interpretation is done using canonical morphisms, which
are just lifting of the same morphisms in $\Rel$, the interpretation of a
proof is the same as its relational interpretation:\footnote{the relational
interpretation is folklore, at least in Marseille and Paris, but it is
surprisingly difficult to find an early reference to it. For those who want to
see the concrete definition of the interpretation, we refer to of~\cite[app.
4]{finiteness}.}
\begin{prop} \label{prop:RelSound}
  For any proof $\pi$ of a sequent $\Gamma|-F$, the relational interpretation
  $[[\pi]]$ of $\pi$ is a simulation from $\bigotimes\Gamma$ to
  $F$. (This holds for any valuation of the propositional variables.)
\end{prop}

\bigbreak
The presence of propositional variables is crucial because without
them, the model becomes trivial:
\begin{prop}
  Suppose $F$ is a formula without propositional variables; then its
  interpretation is trivial : any subset of its set of states is a safety
  property.

  More precisely, we have
  \begin{itemize}
  \item $A_F(s) = \{*\}$ (singleton set) ;
  \item $D_F(s,*) = \{*\}$ (singleton set) ;
  \item $n_F(s,*,*) = s$.
  \end{itemize}
\end{prop}
The proof is a trivial induction on the formula.

This model is thus only appropriate when interpreting $\mathsf{\Pi}^1_1$
logic, \ie propositional linear logic. There, it has a real discriminating
power.
The model can even be extended to deal with full second order, with the usual
proviso: the interpretation may decrease (in some very special cases) during
elimination of a second order cut.  For more details, see~\cite[chap.
8]{PhD}.\footnote{There, the equivalent notion of predicate transformers
rather than interaction systems is used, but as we will show in
section~\ref{sec:PT}, the two categories are equivalent.}


\section{Interpreting the Differential $\lambda$-calculus} 
\label{sec:diff}

So far, proposition~\ref{prop:SupLatticeEnrich} hasn't been used, except to
deduce the existence of a product. The problem is that this proposition
doesn't reflect a property of proofs. The reason is that
\begin{itemize}
\item not every formula has a proof;
\item we do not see \textit{a priori} how to sum proofs of a single formula.
\end{itemize}

Ehrhard and Regnier's \emph{differential $\lambda$-calculus} (\cite{diffLamb})
extends the $\lambda$-calculus by adding a notion of differentiation of
$\lambda$-terms. One consequence is that we need a notion of sum of terms,
interpreted as a non-deterministic choice. It is also possible to only add
sums (and coefficients) to the usual $\lambda$-calculus as
in~\cite{lionelSums}.

It is not the right place to go into the details of the differential
\mbox{$\lambda$-calculus} and we refer to \cite{diffLamb} for motivations and
a complete description.  A complete definition can also be found in the
Appendix on page~\pageref{App:diff}.

\smallbreak
\noindent
In the typed case, we have the following typing rules:
\begin{enumerate}
  \item \infer{}{\Gamma |- 0:\tau}{} and \infer{\Gamma |- t:\tau & \Gamma |- u:\tau}{\Gamma |- t+u:\tau}{};

  \item \infer{\Gamma |- t:\tau->\sigma & \Gamma |- u:\tau}{\Gamma |- \D t·u : \tau->\sigma}{}.
\end{enumerate}
The intuitive meaning is that ``$\D t·u$'' is the result of
(non-deterministically) replacing \emph{exactly one occurrence} of the first
variable of $t$ by $u$. We thus obtain a sum of terms, depending on which
occurrence was replaced.  This gives a notion of differential substitution (or
linear substitution) which yields a \emph{differential-reduction.} The rules
governing this reduction are more complex than usual \mbox{$\beta$-reduction}
rules;  we refer to \cite{diffLamb} or the Appendix.

Besides the natural commutativity and associativity of addition, differential
$\lambda$-terms are also quotiented modulo the following equivalence
relations:
\begin{itemize}
\item $0 = (0)u = \lambda x.0 = \D 0·t = \D t·0$;
\item $(t_1+t_2)\,u = (t_1)u \ +\ (t_2)u$;
\item $\lambda x.(t_1+t_2) = \lambda x.t_1 \ +\ \lambda x.t_2$;
\item $\D (t_1+t_2)·u = \D t_1·u \ +\ \D t_2·u$;
\item $\D t·(u_1+u_2) = \D t·u_1 \ +\ \D t·u_2$;
\item $\D(\D t·u)·v = \D(\D t·v)·u$.
\end{itemize}
The last one is probably the most important one as it allows to link the
notion of differentiation to the traditional, analytic notion of
differentiation. Note that even if the first five rules can be oriented from
left to right, a quotient is inevitable because of the sixth rule and the
commutativity of addition. This quotient is natural because none of those rules
carries any computational content.

\bigbreak
Interpreting usual (without ``$\D$'', ``$+$'' nor ``$0$'') $\lambda$-terms can
be done using the well-known translation of the simply typed
$\lambda$-calculus into intuitionistic linear logic with propositional
variables. Just replace an atomic type by a propositional variable and the
type $\tau->\sigma$ by $!\tau -o \sigma$ inductively. That the resulting
interpretation is sound follows directly from the fact that $\Int$ is a Lafont
category. (In any model for linear logic, the co-Kleisli category of $!\_$ is
cartesian closed.)

The general notion of categorical model for the differential
$\lambda$-calculus (or differential proof nets, or ``differential linear
logic'') is only beginning to emerge. The main paper on the subject
is~\cite{diffCategories}, where the categorical notion of ``differentiation
combinator'' is studied in details. No real soundness theorem is however
proved there because the authors work in a more general setting: the base
category is not necessarily monoidal closed, \ie the co-Kleisli category is
not necessarily cartesian closed.

The notion of differential category is well-suited for our purposes, and we
will show that the category $\Int$ is indeed a differential category. Together
with the fact that~$\Int$ is a Lafont category, it allows to deduce that we do
get a categorical model for the differential $\lambda$-calculus.

\begin{defn}
If $\mathcal{C}$ is a symmetric monoidal category with a coalgebra
modality~$!\_$, we call a natural transformation $d_X : X\Tensor !X -> !X$ a
\emph{deriving transformation} in case the following hold:
\be
d_X\,e_X        &\quad=\quad& 0 \\
d_X\,\Delta     &\quad=\quad& (1\Tensor\Delta)\,(d_X\Tensor 1) \ +\ (1\Tensor\Delta)\,(c\Tensor 1)\,(1\Tensor d_X) \\
d_X\,\epsilon_X &\quad=\quad& (1\Tensor e)\, u_X\\
d_X\,\delta     &\quad=\quad& (1\Tensor\Delta)\,(d_X\Tensor\delta)d_{!X}
\ee
\begingroup\footnotesize
where
\begin{itemize}
\item[-] $e_X:!X->\I$ and $\Delta_X:!X->!X\Tensor !X$ are the operations
  of the coalgebra $!X$;
\item[-] $c:A\Tensor B -> B\Tensor A$ is the symmetry and
  $u_X:X\Tensor \I -> X$ is the unit;
\item[-] $\delta_X:!X->!!X$ and~$\epsilon_X:!X->X$ come from the usual comonad
  laws.\footnote{The applications of the isomorphism $\sigma:(A\Tensor
  B)\Tensor C -> A\Tensor(B\Tensor C)$ are omitted for readability.}
\end{itemize}
\endgroup
\end{defn}
In our case, we can lift $d_X$ from the relational model:
\begin{lem}[and definition]
  For any interaction system $w$ on $S$, the relation $d_w$ defined by
  \[
    d_w \quad=\quad \bigg\{
    \Big((s_0,[s_1,\dots,s_n]),[s_0,s_1,\dots,s_n]\Big)\ |\
    s_0,\dots,s_n \in S\bigg\}
  \]
  is a deriving transformation.
\end{lem}
\begin{pf}
Because it is already shown in~\cite{diffCategories} that this relation is
indeed a deriving transformation in $\Rel$, is suffices to show that $d_w$ is
a simulation from~$w\Tensor!w$ to~$!w$. This is immediate.

\qed\end{pf}
We can now extend the interpretation to differential $\lambda$-terms:
\begin{itemize}
\item the additive structure ($0$ and $+$) is directly interpreted by the monoid
structure ($\emptyset$ and $\cup$);
\item the differential structure is interpreted in the only sensible way:
suppose we have~$\Gamma |- t:\tau->\sigma$ and $\Gamma |- u:\tau$; by
induction, we have $[[t]]\in\Int(!\Gamma,!\tau-o\sigma)$ and
$[[u]]\in\Int(!\Gamma,\tau)$. We can define a morphism in
$\Int\big(!\tau-o\sigma,(\tau\Tensor!\tau)-o\sigma\big)$:
\be
    & 1            &\ \in\ & \Int\big(!\tau-o\sigma\ ,\ !\tau-o\sigma\big)\\
<=> \quad & \widehat1  &\ \in\ & \Int\big((!\tau-o\sigma) \Tensor !\tau\ ,\ \sigma\big)\\
 => \quad & \widehat1\,(1\Tensor d_\tau)
                    &\ \in\ &\Int\big((!\tau-o\sigma) \Tensor \tau\Tensor !\tau\ ,\ \sigma\big)\\
<=> \quad & \widehat{\widehat1\,(1\Tensor d_\tau)}
                   &\ \in\ &\Int\big(!\tau-o\sigma \ ,\ (\tau\Tensor !\tau) -o \sigma\big)
\ee
We write $\D$ for this morphism. This is an internal version of the
differential combinator from~\cite[def. 2.3]{diffCategories}.  From this, we
get
\be
         & \D\,[[t]]           &\ \in\ & \Int\big(!\Gamma\ ,\ (\tau\Tensor!\tau)-o\sigma\big) \\
<=>\quad & \widetilde{\D[[t]]} &\ \in\ & \Int\big(!\Gamma\Tensor\tau\ ,\ !\tau-o\sigma\big)
\ee
which we'll call $[[\D t]]$.  We can now compose interpretations as in:
\[
  !\Gamma \quad\morphism{\Delta}        \quad !\Gamma\Tensor!\Gamma
          \quad\morphism{1\Tensor[[u]]} \quad !\Gamma\Tensor\tau
          \quad\morphism{[[\D t]]}      \quad !\tau-o\sigma
\]
This morphism in $\Int(!\Gamma\ ,\ !\tau-o\sigma)$ is the interpretation of
$\D t·u$.
\end{itemize}
Spelled out concretely in the case of $\Rel$ or $\Int$, the inductive
definition looks like:
\be
  (\gamma,\mu,s') \in [[\D t·u]] \\
  \quad<=>\quad \\
  (\gamma_1,\mu+[s],s') \in [[t]]\ \mbox{for some}\ (\gamma_2,s)\in[[u]] \
  \mbox{s.t.}\ \gamma=\gamma_1+\gamma_2
\ee

\bigbreak
That the interpretation is sound follows rather directly from the properties
of a deriving transformation, see~\cite{diffCategories} for some of the
missing details:
\begin{lem}
For all differential $\lambda$-terms and valuations $\gamma$, we have
\begin{itemize}
\item $[[0]]= [[(0)u]] = [[\lambda x.0]] = [[\D 0·t]]= [[\D t·0]]$;
\item $[[(t_1+t_2)\,u]] = [[(t_1)u \ +\ (t_2)u]]$;
\item $[[\lambda x.(t_1+t_2)]] = [[\lambda x.t_1 \ +\ \lambda x.t_2]]$;
\item $[[\D (t_1+t_2)·u]] = [[\D t_1·u \ +\ \D t_2·u]]$;
\item $[[\D t·(u_1+u_2)]] = [[\D t·u_1 \ +\ \D t·u_2]]$;
\item $[[\D(\D t·u)·v]] = [[\D(\D t·v)·u]]$.
\end{itemize}
\end{lem}
We finally obtain the desired result:
\begin{prop}
  Suppose that $\Gamma |- t:\sigma$ where $\Gamma$ is a context and $t$ a
  differential $\lambda$-term. The relation $[[t]]$ is a simulation relation
  from $!\Gamma$ to~$\sigma$. Moreover, for all $t$ and $u$ we have:
  \be
    [[ (\lambda x.t)u ]]    &=& [[ t[u/x] ]] \cr
    [[ \D(\lambda x.t)·u ]] &=& [[ \lambda x \ .\  (\partial t/\partial x)·u ]]
  \ee
\end{prop}
\begin{pf}
  That we obtain a simulation is true by construction.

  Invariance under $\beta$-reduction follows from the correctness of the
  interpretation of $\lambda$-calculus in a cartesian-closed category.

  Invariance under linear substitution seems to follow from general
  considerations about deriving transformation in linear categories, even if
  this is not treated in~\cite{diffCategories}. (In our case, a direct
  verification is possible, but long and tedious...)

\qed\end{pf}

\section{Untyped Calculus} 

Interpreting \emph{untyped} $\lambda$-calculus remained an open question for
quite a long time: since the cardinality of a function space is strictly
bigger than the cardinality of the original set, it seemed difficult to get a
model where any $\lambda$-term can be either an argument or a function. Dana
Scott finally found a model by constructing a special object in the category
of domains.

The solution is quite elegant: to interpret untyped $\lambda$-terms in a
cartesian closed category, one ``just'' needs to find a reflexive object in a
cartesian closed category, \ie an object $X$ with a retraction / projection
pair $[X->X] \triangleleft X$.

We have at our disposal a cartesian closed category: the Kleisli category over
the comonad $!\_$. Were we to find a reflexive object $W$ in this category, we
could model the \emph{untyped differential} $\lambda$-calculus in the
$\lambda$-model $\Saf(W)$.
We now show how to construct such a reflexive object, in a fairly
straightforward way.


\smallbreak
We start with a non-trivial interaction system $w$ on a set of states
$S$ (natural numbers for example) and then define an interaction system
$W$, satisfying the equation $W \simeq w \Plus (!W-oW)$ as follows:
\begin{enumerate}
\item the set of states $S_W$ is defined as the least fixpoint of $X|->S +
\Mulf(X)×X$; in a more ``programming'' fashion''
\be
  S_W \qquad=\qquad  \textbf{data} &   &\mathsf{Leaf}(s\in S) \\
                                   &|\ & \mathsf{Node}\Big(\mu\in\Mulf(S_W)\,,\,u\in S_W\Big)
\ee

\item the possible actions in a given state are defined by induction on the
state: using ``pattern matching'', we have
\be
  A_W(\mathsf{Leaf}(s))     &\quad=\quad& A(s)  \\
  A_W(\mathsf{Node}(\mu,u)) &\quad=\quad& (!A_W -o A_W)((\mu,u))
\ee

\item reactions are defined similarly as
\be
  D_W(\mathsf{Leaf}(s),a)     &\quad=\quad& D(s,a)  \\
  D_W(\mathsf{Node}(\mu,u),b) &\quad=\quad& (!D_W -o D_W)((\mu,u),b)
\ee

\item and finally, the next state function is defined as
\be
  n_W(\mathsf{Leaf}(s),a,d)     &\quad=\quad& \mathsf{Leaf}(s[a/d])  \\
  n_W(\mathsf{Node}(\mu,u),b,p) &\quad=\quad& (!n_W -o n_W)((\mu,u),b,p)
\ee
\end{enumerate}
Due to the presence of multisets, an actual implementation of $W$ in
a dependently typed functional programming language would be a little more
complex: we would need to work with lists rather than multisets, and reason
modulo shuffling concretely.

\medbreak
\begin{lem}
  The relation $r$ between $S + (\Mulf(S_W)×S_W)$ and $S_W$ defined by
  \be
    \big((1,s)\,,\,\mathsf{Leaf}(s)\big) &\quad\in\quad& r\\
    \big((2,(\mu,u))\,,\,\mathsf{Node}(\mu,u)\big) &\quad\in\quad& r
  \ee
  is an isomorphism (in $\Int$) from $w\Plus (!W-oW)$ to $W$.
\end{lem}
The proof is direct. (This is an instance of a strong, ``componentwise''
isomorphism.)

\begin{cor}
  In the $!\_$-Kleisli category of $\Int$, which is cartesian closed, there is
  a retract $W^W\triangleleft W$.
\end{cor}
\begin{pf}
First, notice that it is sufficient to find a retract in the category~$\Int$:
any morphism in a category can also be seen as a morphism in a Kleisli
category (in a way which is compatible with composition in the Kleisli
category).

There is the canonical injection $i_2$ from $!W-oW$ to $w\Plus (!W{-o}W)$.
Now, in the category~$\Int$, we have that product and coproduct coincide; in
particular, we have the projection $\pi_2$ from $w\With(!W{-o}W) =
w\Plus(!W{-o}W)$ to $!W-oW$.  Moreover, by definition, we have
$\pi_2·i_2=\Id_{!W-oW}$.

We can now prove that $r·i_2$ / $\pi_2·r^\sim$ is a retraction / projection
from~$!W-oW$ to~$W$:\footnote{\label{converse}The converse $r^\sim$ of a
relation is defined as $(s_2,s_1)\in R^\sim$ iff $(s_1,s_2)\in r$.}
it follows from the previous remark that $\pi_2·i_2=\Id$ and that $r^\sim$ is
the inverse of $r$. (Recall that $r$ is an isomorphism.)
\qed\end{pf}

From there, constructing a model for the untyped $\lambda$-calculus is
standard. We refer to~\cite{Curien}. We obtain in this way a model where each
term is interpreted by a safety property for $W$.

Since $\Saf(W)$ is a complete sup-lattice, we can also model sums, and by the
very same construction defined in section~\ref{sec:diff}, model
differentiation. (Remark that the interpretation of a term is not really by
induction on the type inference, but directly by induction on the term: we can
thus apply it to untyped terms as well.)

\medbreak
The interpretation becomes: if $t$ is a differential $\lambda$-term with its
free variables among $x_1,\dots,x_n$, we interpret $t$ by a subset of
$\Mulf(S_W)×\cdots×\Mulf(S_W)×S_W$. In the sequel, $\gamma$ is a tuple in
$\Mulf(S_W)×\cdots×\Mulf(S_W)$ and we use $\gamma(x)$ for the projection on
the appropriate coordinate.
\begin{itemize}

\item
    $[[x]] = \{(\gamma,s)\}$ where $\gamma(x)=[s]$ and $\gamma(y)=[]$
    otherwise

\smallbreak
\item
    $[[\lambda x.t]] = \big\{ \big(\gamma,(\mu,s)\big) || 
    (\gamma_{x:=\mu},s)\in[[t]]\big\}$

\smallbreak
\item
    $(\gamma,s)\in[[(t)u]]$ iff
    $\left(\begin{tabular}[c]{l}
    $\big(\gamma_0,\mathsf{Node}(\mu,s)\big)\in[[t]]$ for some
    $\mu=[s_1,\ldots,s_n]$ \\
    s.t.  $(\gamma_i,s_i)\in[[u]]$ for $i=1,\ldots,n$\\
    and $\gamma=\gamma_0+\gamma_1+\ldots+\gamma_n$;
    \end{tabular}\right.$

\smallbreak
\item
    $[[0]] = \emptyset$;

\smallbreak
\item
    $[[t_1+t_2]] = [[t_1]] \cup [[t_2]]$;

\smallbreak
\item
    $(\gamma,\mu,s') \in [[\D t·u]]$ iff
     $\left(\begin{tabular}[c]{l}
     $\big(\gamma_1,\mathsf{Node}(\mu+[s],s')\big) \in [[t]]$\\
     for some $(\gamma_2,s)\in[[u]]$
     s.t. $\gamma=\gamma_1+\gamma_2$.
    \end{tabular}\right.$

\end{itemize}

\begin{prop}
  For any closed differential $\lambda$-term $t$, we have that $[[t]]$ is a
  safety property for $W$.
\end{prop}
We have thus, in effect, constructed a non-trivial (in the sense that not
\emph{all} subsets of $S_W$ are safety properties) denotational model for the
untyped differential $\lambda$-calculus. This is particularly interesting
because the original model for differential $\lambda$-calculus (finiteness
spaces) did not have a reflexive object: they could not interpret fixpoint
combinators (see~\cite{finiteness}).

\section{Classical Linear Logic} 

If one has in mind the definition of negation (see page~\pageref{def:Bottom}),
the next result can look quite surprising: interaction systems can interpret
\emph{classical} linear logic. In other words, for any interaction system $w$,
we have $w\simeq w^{\Bottom\Bottom}$. The reason behind that is that our
notion of morphism is \emph{not} the notion of ``componentwise'' morphism.
Even though the actions/reactions in $w^{\Bottom\Bottom}$ are very complex
sets, the way they interact with states remains relatively simple.

The reason we haven't shown this result in section~\ref{sec:ILL} is that the
principle at work is highly non-constructive and that natural generalizations
of interaction systems are unlikely to satisfy it.

Recall that any object can be used to represent $\Bottom$ in the
intuitionistic case. However, in or case, the object $\I$ plays a very special
role. For $\Bottom=\I$, we have:
\begin{prop} \label{prop:dual}
  In $\Int$, for any interaction system $w$, the identity relation is an
  isomorphism between $w$ and $w^{\Bottom\Bottom}$.

  Equivalently, the object $\Bottom$ is dualizing in $\Int$.
\end{prop}
\begin{pf}
The principle at stake in the proof is the contrapositive of the axiom
of choice:
\be
\CtrAC :\quad &
\big(\exists a\in A\big)\big(\forall d\in D(a)\big)\ \varphi(a,d) \\
           &<=>\\
&\big(\forall f\in\prod_{a\in A} D(a)\big)\big(\exists a\in A\big)\ \varphi\big(a,f(a)\big)
\ee
When the domain $D(a)$ for the universal quantifier doesn't depend on
$a\in A$, we can simplify it into:
\be
\CtrAC :\quad &
\big(\exists a\in A\big)\big(\forall d\in D\big)\ \varphi(a,d) \\
 & <=> \\
& \big(\forall f\in A -> D\big)\big(\exists a\in A\big)\ \varphi\big(a,f(a)\big)
\ee
Here are the components of $w^{\Bottom\Bottom}$:
\be
  A^{\Bottom\Bottom}(s)     &\quad=\quad& \displaystyle\left(\prod_{a\in A(s)} D(s,a) \right) -> A(s) \\
  D^{\Bottom\Bottom}(s,F)   &\quad=\quad& \displaystyle\prod_{a\in A(s)} D(s,a)     \\
  n^{\Bottom\Bottom}(s,F,g) &\quad=\quad& s\big[F(g)/g\big(F(g)\big)\big]
  \ \mbox{.}
\ee
That equality is a simulation from $w^{\Bottom\Bottom}$ to $w$ takes the form:
\be
  (\forall s\in S)\quad & \big(\forall F\in A^{\Bottom\Bottom}(s)\big)
                          \bm{\big(\exists a\in A(s)\big)}\\
                        & \bm{\big(\forall d\in D(s,a)\big)}
                          \big(\exists g\in D^{\Bottom\Bottom}(s,F)\big)\\
                        & \qquad s[a/\bm{d}] =_S s\big[F(g)/g\big(F(g)\big)\big]
\ \mbox{.}
\ee
By applying the contraposition of the axiom of choice
on $\exists a\forall d$, this is equivalent to
\be
  (\forall s\in S)\quad & \big(\forall F\in A^{\Bottom\Bottom}(s)\big)
                          \left(\forall f\in \prod_{a\in A(s)} D(s,a)\right)\\
                        & \big(\exists a\in A(s)\big)
                          \big(\exists g\in D^{\Bottom\Bottom}(s,F)\big)\\
                        & \qquad s[a/d] =_S s\big[F(g)/g\big(F(g)\big)\big]
\ \mbox{.}
\ee
We can swap quantifiers and obtain, by the definitions of $A^\Bottom$,
$D^\Bottom$ and $A^{\Bottom\Bottom}$,
\be
  (\forall s\in S)\quad & \big(\forall f\in A^\Bottom(s)\big)
                          \bm{\big(\forall F\in A^\Bottom(s) \to D^\Bottom(s,\_)\big)} \\
                        & \bm{\big(\exists g\in D^{\Bottom\Bottom}(s,F)\big)}
                          \big(\exists a\in D^\Bottom(s,f)\big)\\
                        & \qquad s[a/f(a)] =_S s\big[\bm{F(g)}/g\big(\bm{F(g)}\big)\big]
\ \mbox{.}
\ee
We can now apply the contraposition of the axiom of choice on $\forall
F\exists g$ to get the equivalent formulation
\be
  (\forall s\in S)\quad & \big(\forall f\in A^\Bottom(s)\big)
                          \big(\exists g\in D^{\Bottom\Bottom}(s,F)\big) \\
                        & \big(\forall b\in D^{\Bottom}(s,g)\big)
                          \big(\exists a\in D^\Bottom(s,f)\big)\\
                        & \qquad s[a/f(a)] =_S s[b/g(b)]
\ \mbox{.}
\ee
Since $D^{\Bottom\Bottom}$ is equal to $A^\Bottom$, this is obviously true.

\smallbreak
Thus, we can conclude that equality is a simulation from $w^{\Bottom\Bottom}$ to
$w$.
\qed\end{pf}
We obtain a surprising corollary:
\begin{cor}  \label{cor:ISSimple}
  Any interaction system is isomorphic to an interaction where the sets of
  reactions do not depend on a particular action. (More precisely, for any
  state $s$, the function $a |-> D(s,a)$ is constant.)
\end{cor}
\begin{pf}
Just notice that $w^{\Bottom\Bottom}$ satisfies this property.
\qed\end{pf}

Finally, we have
\begin{cor}
  The category $\Int$ is $\star$-autonomous (see~\cite{barr}), we can thus
  interpret classical linear logic.
\end{cor}
\begin{pf}
Once we know that $\Bottom$ is dualizing, the remaining condition are fairly
easy to check: the following diagram should be commutative
\diagram[height=3em,width=5em]
 w_1 -o w_2                   & \rTo^{\_^\Bottom}       & w_2^\Bottom -o w_1^\Bottom \\
 &\rdTo_{d_{w_1}^{-1}·\_·d_{w_2}}    & \dTo_{\_^\Bottom} \\
             && w_1^{\Bottom\Bottom} -o w_2^{\Bottom\Bottom}               \\
\enddiagram
where $d_w$ is the natural isomorphism from $w$ to $w^{\Bottom\Bottom}$. This
is immediate since~$d_w$ is the identity on $S$ and $\_^\Bottom$ is the
``converse'' operation of a relation. (See footnote~\ref{converse} on
page~\pageref{converse}.)

We can then unfold all the usual technology to give a denotational
model for \emph{classical} linear logic.

\qed\end{pf}

It is interesting to highlight some aspects of this model
\begin{itemize}
\item $\Int$ is a ``games'' model for full \emph{classical} linear logic. The
isomorphism between $w$ and $w^{\Bottom\Bottom}$ is given by the identity
relation.
\item The constructions are quite different from usual games constructions; in
particular, they have a strong \emph{synchronous} feeling.
\item The notion of strategy is not used to define morphisms; rather, we use
the notion of \emph{simulation.}
\item The fact that $w\simeq w^{\Bottom\Bottom}$ seems rather accidental as it
is not expected to hold in any generalized version of interaction systems.
(See the discussion about containers in section~\ref{sec:Containers}). This
fact is also highly non-constructive and almost counter-intuitive.
\end{itemize}

Putting the model of the differential $\lambda$-calculus with the dualizing
object $\Bottom$, it is expected that we get a model for Lionel Vaux's
``differential $\lambda\mu$-calculus'' (see~\cite{lionelDiffLamb}), either in
Vaux's setting (typed) or in an untyped setting.

\section{Related Notions} 

\subsection{Predicate Transformers} 
\label{sec:PT}

The category $\Int$ has a very concrete feeling. In \cite{denotPT}, we have
developed a model for full linear logic with a different intuition: the
category of \emph{predicate transformers} with \emph{forward data-refinements:}
\begin{defn}
  If $S$ is a set, a \emph{predicate transformer} on $S$ is a monotonic
  (w.r.t. inclusion of subsets) function from $\Pow(S)$ to $\Pow(S)$.

  If $F_1$ and $F_2$ are predicate transformers respectively on $S_1$ and $S_2$, a
  \emph{forward data-refinement} from $P_1$ to $P_2$ is a relation $r\sub
  S_1×S_2$ s.t. $\langle r\rangle · F_1 \sub F_2·\langle
  r\rangle$.\footnote{Where $\langle r\rangle$ is the \emph{direct image} of
  $r$: $s_2\in\langle r\rangle(x)$ iff $(\exists s_1)\,(s_1,s_2)\in r
  \land s_1\in x$.}
  (Extensional ordering.)

  Such predicate transformers with forward data-refinements form a category
  called~$\PT$.
\end{defn}
An interaction system can be seen as a concrete representation for a predicate
transformer. More precisely:
\begin{prop}  \label{prop:IntPT}
  The operation $w|->w^\circ$ from $\Int$ to $\PT$ defined as
  \[
    s \in w^\circ(x) \quad<=>\quad (\exists a\in w.a(s))(\forall d\in
    w.D(s,a))\ s[a/d]\in x
  \]
  can be extended to a full and faithful functor from $\Int$ to $\PT$.
\end{prop}
The intuition in $s\in w^\circ(x)$ is that the Angel has a foolproof way to
reach $x$ in exactly one interaction.
\begin{pf}
  The action on morphisms is just the identity: we thus need to show that $r$
  is a simulation from $w_1$ to $w_2$ iff $r$ is a forward data-refinement
  from~$w_1^\circ$ to~$w_2^\circ$. The proof is not very difficult and can be
  found in~\cite{PhD}.
\qed\end{pf}
The interesting point is that all the constructions presented above are the
concrete versions of the constructions presented in~\cite{denotPT}. For example,
we have
\[
  (w_1\Tensor w_2)^\circ \quad=\quad w_1^\circ \Tensor w_2^\circ
\]
where $\Tensor$ on the left is the synchronous tensor on interaction systems
and the tensor on the right is the tensor on predicate
transformers.\footnote{$(s_1,s_2) \in F_1{\Tensor}
F_2(r)$ iff $(\exists x{\sub}S_1)(\exists y{\sub}S_2)\, x{×}y\sub r \land s_1\in
F_1(x)\land s_2\in F_2(y)$}

Another interesting example is the fact that $(w^\Bottom)^\circ =
(w^\circ)^\Bottom$. This is interesting because the definition of duality in
the case of predicate transformers is very simple, and involutivity is trivial:
\[
  F^\Bottom(x) \quad=\quad \overline{F(\overline{x})}
\]
where $\overline{x}$ represents the $S$-complement of $x$ (\ie
$\overline{x}=S\setminus x$).

\bigbreak
Surprisingly, proposition~\ref{prop:IntPT} can be strengthened in an ad-hoc way
to read:
\begin{prop} \label{prop:PTInt}
  The categories $\Int$ and $\PT$ are equivalent. Moreover, this equivalence
  is a ``retract''.
\end{prop}
\begin{pf}
By ``retract'', we mean the following: there is a functor $\mathcal{F}$ from
$\PT$ to $\Int$ which satisfies $\mathcal{F}(F)^\circ = F$ and
$\mathcal{F}(w^\circ)\simeq w$. In other words, we obtain equal object in one
direction but only isomorphic objects in the other direction.

This functor $\mathcal{F}$ is defined as follows: let $F$ be a predicate
transformer on $S$, define $\mathcal{F}$ to be the interaction system on $S$
with components
\be
  \mathcal{F}(F).A(s)      &\quad=\quad& \{ x\sub S\ \mid\ s\in F(x)\} \\
  \mathcal{F}(F).D(s,x)    &\quad=\quad& x \\
  \mathcal{F}(F).n(s,x,s') &\quad=\quad& s'
\ee
Checking that this operation does define a functor is left as an exercise.
(See~\cite{PhD}.)
\qed\end{pf}
Once more, we separate propositions~\ref{prop:IntPT} and~\ref{prop:PTInt}
because while we expect proposition~\ref{prop:IntPT} to hold for different
generalizations of interaction systems / predicate transformers (see
theorem~3.4 in~\cite{cont}), proposition~\ref{prop:PTInt} seems very specific
to this particular case.

\subsection{Containers}
\label{sec:Containers}

In \cite{cont}, the authors study the notion of \emph{container,} a structure
bearing several similarities with the notion of interaction system.  They work
in a variant of Martin-L\"of type theory (\cite{ML84,NPS}), a dependent
predicative type theory.

Mild knowledge about this type theory is assumed in this section.

Simply, a container is given by the following:
\begin{itemize}
\item a set $A$ of \emph{shapes};
\item and for any $a\in A$, a set $D(a)$ of \emph{positions.}
\end{itemize}
A morphisms from $(A_1,D_1)$ to $(A_2,D_2)$ is given by a pair
$(f,u)$ where $f$ is a function $f:A_1->A_2$ and $u$ is a family of functions
indexed by $A_1$ and we have $u_{a_1} : D_2(f(a_1)) -> D_1(a_1)$.

This is reminiscent of interaction systems in the following way: any
interaction system on the set of states $\{*\}$ (singleton set) can be seen as
a container, and any container can be seen as an interaction system on
$\{*\}$.

The links between container morphisms and simulations is subtler: a simulation
from $w_1$ to $w_2$ (two interaction systems on $\{*\}$) is given by a
relation~$r\sub \{*\}×\{*\}\simeq\{*\}$. In Martin-L\"of type theory, a subset
is seen as a propositional function $r:\{*\}->\Set$, \ie a set.  The condition
required to make this ``relation'' a simulation is the following:
\be
  r &\qquad=>\qquad& (\forall a_1\in A_1)(\exists a_2\in A_2) \\
    &              & \big(\forall d_2\in D_2(a_2)\big)\big(\exists d_1\in D_1(a_1)\big)\ r
  \ \mbox{.}
\ee
We can apply the (constructive) axiom of choice to skolemize this and we
obtain
\be
  r &\qquad=>\qquad& (\exists f:A_1->A_2)
                     \Big(\exists u:\prod_{a_1\in A_1} D_2\big(f(a_1)\big) -> D_1(a_1)\Big) \\
    &              & (\forall a_1\in A_1)\big(\forall d_2\in D_2(f(a_1))\big)\ r
\ee
which is \emph{logically} equivalent to
\be
  r \qquad=>\qquad (\exists f:A_1->A_2)
                   \Big(\exists u:\prod_{a_1\in A_1} D_2\big(f(a_1)\big) -> D_1(a_1)\Big)
                   \ \top
\ee
where $\top$ denotes the true proposition (or the singleton set in Martin
L\"of type theory).

Thus, a constructive (in the sense of Martin-L\"of type theory) simulations
between two interaction systems on $\{*\}$ is given by:
\begin{itemize}
\item a set $r$;
\item and a function $r -> (\Sigma f:A_1->A_2)\,\prod_{a_1\in A_1}
D_2\big(f(a_1)\big) -> D_1(a_1)$.
\end{itemize}
Equivalently, a simulation between two interaction systems on $\{*\}$ is
nothing but a \emph{family of container morphisms} between the corresponding
containers!

However, the difference is that while container morphisms are identified when
the \emph{functions} acting on actions / reactions are extensionally equal,
simulations are identified when the \emph{relations} between states are
extensionally equal. In other words, two simulations $(r_1,(f_1,u_1))$ and
$(r_2,(f_2,u_2))$ between interaction systems on $\{*\}$ are equal when there
are ``translating functions'' $r_1
\mathrel{\lower3pt\hbox{$\mathop{\stackrel{\displaystyle\rightarrow}{\leftarrow}}$}}
r_2$.\footnote{Note that since $r_1$ and $r_2$ are ``propositions'', we do not
require that those translating functions are inverse of each other.}

\medbreak
If we adopt a classical point of view, then everything is rather boring:
there is at most one non-empty simulation between two interaction systems on
$\{*\}$: the relation $\{(*,*)\}$. The links with container morphisms is then
the following:
\begin{quote}\sl
  if there is at least one container morphisms from $w_1$ to $w_2$, then there
  will be exactly one non-empty simulation from $w_2$ to $w_2$;\\
  if there are no container morphism from $w_1$ to $w_2$, then the only
  simulation from $w_1$ to $w_2$ will be the empty simulation.
\end{quote}

\medbreak
This whole theory of containers can be extended to work in a large class of
locally cartesian closed categories. (See~\cite{cont}.) In such a setting, one
needs to take care of additional coherence diagrams, but the idea is
similar.

There is currently some work being done on generalizing containers to a notion
of dependent containers, \ie interaction systems.  The idea, following the
original intuition of~\cite{Seely} and~\cite{Hofmann} is to define a dependent
container in a locally cartesian closed category $\mathcal{C}$ as:
\begin{itemize}
\item an object $S$ in $\mathcal{C}$;
\item an object $A$ in $\mathcal{C}/S$;
\item an object $D$ in $\mathcal{C}/(\Sigma_S A)$;\footnote{Recall that in a
locally cartesian closed category, if $B$ is a slice in $\mathcal{C}/A$, we
write~$\Sigma_A B$ for the codomain of $B$.}
\item a morphism $n$ in $\mathcal{C}(\Sigma_{\Sigma_S A} D \,,\,S)$.
\end{itemize}
The appropriate notion(s) of morphism is not entirely clear and still under
heavy discussion...

\section*{Conclusion} 

We have developed a new category where objects are games to model linear
logic or $\lambda$-calculus. What is rather new is the use of a notion of
\emph{simulation} for morphisms: the notion of strategy is not used in this
model! Strategies do however appear implicitly in the notion of safety
property which are the ``points'' of our model: a safety property is set of
states for which there is an infinite strategy which restrict interaction to
stay in the safety property. This strategy is only guaranteed to exists, but
there is in general no way to obtain it.

Some of the interesting points about this model are that is allows to model
full linear logic (it can even be extended to second order). Moreover, and
this is relatively new, it can interpret the untyped differential
$\lambda$-calculus.

An interesting project is to see whether one can apply the technology
developed here in order to give denotational models for more interesting
programming languages. PCF-like languages ought to be rather easy, but
interaction systems (or predicate transformers) are rooted in ``real''
programming\footnote{Predicate transformers were used to give a semantics to
sequential programs by Dijkstra (\textbf{wp} or \textbf{wlp} calculus) and
have been extended to deal with \emph{specifications} as well (whole field of
refinement calculus); interaction systems, as hinted in the first section seem
appropriate to describe interfaces.} so that we might expect to have access to
many programming features...

\appendix

\section{The Simply Typed Differential $\lambda$-calculus}  
\label{App:diff}

The syntax of the simply typed differential $\lambda$-calculus is given by the
following grammar:
\be
  t,u,t_1,t_2 &\qquad::=\qquad&
     x                \quad|\quad
     (t_1)\,t_2       \quad|\quad
     \lambda x.t      \quad|\quad \\
     && 0                \quad|\quad
     (t_1 + t_2)      \quad|\quad
     \D t·u
\ee
We define $\beta$-reduction in the usual way:
\[
  (\lambda x.t)\,u \quad \leadsto_\beta \quad t[u/x]
\]
where substitution is extended in the following ``obvious'' way:
\be
  x[u/x]             &\quad=\quad&  u \\
  y[u/x]             &\quad=\quad&  y \quad\mbox{if $y\neq x$} \\
  (t)v\ [u/x]        &\quad=\quad&  (t[u/x]) v[u/x]\\
  \lambda x.t\ [u/x] &\quad=\quad&  \lambda x.t \\
  \lambda y.t\ [u/x] &\quad=\quad&  \lambda y\ .\ t[u/x] \quad\mbox{if $y\neq x$} \\
  0[u/x]             &\quad=\quad&  0\\
  t_1+t_2\ [u/x]     &\quad=\quad&  t_1[u/x]\ +\ t_2[u/x]\\
  \D t·v\ [u/x]      &\quad=\quad&  \D t[u/x]\ ·\ v[u/x]
\ee

\emph{Differential reduction} is defined as:
\[
  \D (\lambda x.t)·u \quad\leadsto_{\D}\quad \lambda x \ .\  \frac{\partial
  t}{\partial x}·u
\]
where $\partial t/\partial x·u$, the \emph{linear substitution of $x$ by $u$
in $t$} is defined as:
\be
  \partial x/\partial x\ ·\ u
    &\quad=\quad& u \cr
  \partial y/\partial x\ ·\ u
    &\quad=\quad& 0 \quad\mbox{if $y\neq x$}\cr
  \partial (t)v/\partial x\ ·\ u
    &\quad=\quad&
      (\partial t/\partial x·u) v + \big(\D t·(\partial v/\partial x·u)\big) v \cr
  \partial \lambda x.t/\partial x \ ·\ u
    &\quad=\quad& \lambda x.t \cr
  \partial \lambda y.t/\partial x\ ·\ u
    &\quad=\quad&
      \lambda y.(\partial t/\partial x·u)\quad\mbox{if $y\neq x$} \cr
  \partial 0/\partial x\ ·\ u
    &\quad=\quad& 0\cr
  \partial (t_1+t_2)/\partial x\ ·\ u
    &\quad=\quad&
      \partial t_1/\partial x·u \ +\  \partial t_2/\partial x·u\cr
  \partial (\D t·v)/\partial x\ ·\ u
    &\quad=\quad&
      \D (\partial t/\partial x·u)·v \ +\  \D t·(\partial v/\partial x·u)
\ee
Terms are quotiented by the (contextual closure of the) following equations:
\begin{itemize}
\item $0 = (0)u = \lambda x.0 = \D 0·t = \D t·0$;
\item $(t_1+t_2)\,u = (t_1)u \ +\ (t_2)u$;
\item $\lambda x.(t_1+t_2) = \lambda x.t_1 \ +\ \lambda x.t_2$;
\item $\D (t_1+t_2)·u = \D t_1·u \ +\ \D t_2·u$;
\item $\D t·(u_1+u_2) = \D t·u_1 \ +\ \D t·u_2$;
\item $\D(\D t·u)·v = \D(\D t·v)·u$.
\end{itemize}

The typing rules are
\begin{enumerate}
  \item \infer{}{\Gamma |- x:\tau}{} if ``$x:\tau$'' appears in $\Gamma$;
  \item \infer{\Gamma,x:\tau |- t:\sigma}{\Gamma |- \lambda x.t :
  \sigma->\tau}{};
  \item \infer{\Gamma |- t:\tau -> \sigma & \Gamma |- u:\sigma}{\Gamma |-
  (t)\,u : \tau}{};
  \item \infer{}{\Gamma |- 0:\tau}{} and \infer{\Gamma |- t:\tau & \Gamma |- u:\tau}{\Gamma |- t+u:\tau}{};

  \item \infer{\Gamma |- t:\tau->\sigma & \Gamma |- u:\tau}{\Gamma |- \D t·u : \tau->\sigma}{}.
\end{enumerate}
Typed terms enjoy the Church-Rosser property and strong normalization (w.r.t.
$\beta$/differential reduction).


\bibliographystyle{elsart-num}
\bibliography{interaction}


\end{document}